\documentclass{ws-procs975x65}

\def\beq{\begin{equation}}
\def\eeq{\end{equation}}

\begin{document}

\title{Impacts of photon bending on observational aspects of Two Component Advective Flow}
\author{$Arka$~$Chatterjee^{1^*}$ and $Sandip$~$K.$~$Chakrabarti^{1,2}$}

\address{1. Indian Centre for Space Physics, Chalantika 43\\
Garia Station Road, Kolkata, India, $email: arka@csp.res.in$}
\address{2. S. N. Bose National Centre for Basic Sciences, Block-JD\\ 
Salt Lake, Kolkata, India, $email: chakraba@bose.res.in$}

\begin{abstract}
Nature of photon trajectories in a curved spacetime around black holes are studied 
without constraining their motion to any plane. Impacts of photon bending are 
separately scrutinized for Keplerian and CENBOL components of Two Component 
Advective Flow (TCAF) model. Parameters like Red shift, Bolometric Flux, 
temperature profile and time of arrival of photons are also computed.

\end{abstract}

\keywords{Black Hole; Null Geodesics; Redshift; Accretion disk}

\bodymatter


\section{Introduction}

It is a well known that hard X-rays originate from regions very close to black holes
(Chakrabarti \& Titarchuk, 1995, hereafter CT95). Thus, the effects of photon bending 
(Luminet, 1979; Weinberg, 1972) are of major importance. Furthermore, a power-law tail 
produced by the bulk motion Comptonization from regions between the inner sonic point
and the horizon is expected to be dominated by such effects. We solve 3D photon 
trajectory equations (Weinberg 1972) to trace the path of each ray from the emission 
region to the observer plane. In CT95, major components are the geometrically thin 
Keplerian disk in the pre-shock flow and geometrically thick torus or CENBOL is the 
post-shock region. Images of optically thick Keplerian disk and thick accretion disk 
(Chakrabarti, 1985) are developed separately. In future, these will be merged and
the composite spectra will be showed.

\subsection{Photon Trajectory Equations}
In four dimensions, the motion of free particles or photons are governed by the
following equation,
\noindent
\begin{equation}
\frac{d^2x^{\mu}}{dp^2}+ {\Gamma}_{\nu\lambda}^{\mu}\frac{dx^{\nu}}{dp}\frac{dx^{\lambda}}{dp} = 0,
\end{equation}
with ${\mu} = [0,1,2,3]$; $x^{0} = t$, $x^{1} = r$, $x^{2} = \theta$ and $x^{3} = \phi$,
where $p$ is our affine parameter and ${\Gamma}_{\nu\lambda}^{\mu}$ is Christoffel symbols.

Expanding equation $(1)$, we get $1+3$ coupled differential equations
of second order. Specific energy $P_t=E=(1-\frac{1}{r})\frac{dt}{dp}$ and specific angular
momentum $P_{\phi}=L=r^{2}sin^{2}\theta\frac{d\phi}{dp}$ are two physical quantities
which characterizes a trajectory (Chandrasekhar 1983). Since, the path
of a photon are is guided by it's energy, we can consider $P_t=E=1$ which will be 
modified later by flux equation. However, $P_{\phi}=L$ is a variable which 
changes impact parameters as defined by $b=L/E$ (Luminet 1979).

This condition allows us to drop one equation by substituting
the derivative from affine parameter $(p)$ to time $(t)$ co-ordinate. Thus we get
in Schwarzschild geometry as:
\begin{equation}
\begin{aligned}
\frac{d^2r}{dt^2} + \frac{3}{2}\frac{1}{r(r-1)}\bigg(\frac{dr}{dt}\bigg)^2 - 
(r-1)\bigg(\frac{d\theta}{dt}\bigg)^2 - (r-1)r\mathrm{sin}^2{\theta}\bigg(\frac{d\phi}{dt}\bigg)^2 
+ \frac{r-1}{2r^3} = 0,\\
\frac{d^2\theta}{dt^2} + \frac{2r-3}{r(r-1)}\bigg(\frac{d\theta}{dt}\bigg)\bigg(\frac{dr}{dt}\bigg) 
- \mathrm{sin}{\theta}\mathrm{cos{\theta}}\bigg(\frac{d\phi}{dt}\bigg)^2 = 0 ~\mathrm{and}\\
\frac{d^2\phi}{dt^2} +\frac{2r-3}{r(r-1)}\bigg(\frac{d\theta}{dt}\bigg)\bigg(\frac{d\phi}{dt}\bigg)
 + 2\mathrm{cot}{\theta}\bigg(\frac{d\theta}{dt}\bigg)\bigg(\frac{d\phi}{dt}\bigg) = 0.
\end{aligned}
\label{eq:xdef}
\end{equation}
We have three second order coupled differential equations which govern
the path of a photon in this curved geometry. Three spatial velocity components 
are
\begin{equation}
\begin{aligned}
v^{\hat{r}}=\frac{d\hat{r}}{dt}=\frac{r}{(r-1)}\frac{dr}{dt},~v^{\hat{\theta}}=\frac{d\hat{\theta}}{dt}=\frac{r\sqrt{r}}{\sqrt{(r-1)}}\frac{d\theta}{dt}\\ i\mathrm{and} ~
v^{\hat{\phi}}=\frac{d\hat{\phi}}{dt}=\frac{r\sqrt{r}\mathrm{sin}{\theta}}{\sqrt{(r-1)}}\frac{d\theta}{dt}.
\end{aligned}
\label{eq:xdef}
\end{equation}
The set of six ($r_{\circ}$, $\theta_{\circ}$, $\phi_{\circ}$, $v^{\hat{r}}_{\circ}$, 
$v^{\hat{\theta}}_{\circ}$ and $v^{\hat{\phi}}_{\circ}$) initial
conditions are supplied to solve this set of equations.

\section{Redshifts of Photons}
To generate spectrum and image one must introduce the concept of redshift.
There are two contributors of the redshift. One comes from the Doppler 
effect caused by rotational motion of the disk and another is gravitational 
redshift caused by the black hole. So, redshift can be written as:
\begin{equation}
1+z= z_{grav}z_{rot},
\end{equation}
where, $z_{grav}$ and $z_{rot}$ are gravitational and rotational contributions. 
By definition, redshift z is:
\begin{equation}
1+z= \frac{E_{em}}{E_{obs}}=\frac{(P_{\alpha}u^{\alpha})^{em}}{(P_{\alpha}u^{\alpha})^{obs}},
\end{equation}
where, $E_{em}$ and $E_{obs}$ are energy of emitted and observed photons.
Taking and inner product of 4-momentum and 4-velocity gives the energy of emitted
photon as
\begin{equation}
 E_{em}=P_tu^t+P_{\phi}u^{\phi}=P_tu^t\bigg(1+\Omega_k\frac{P_{\phi}}{P_t}\bigg),
\end{equation}
where $\Omega_k=\frac{u^{\phi}}{u^t}$. From earlier definition, $\frac{P_{\phi}}{P_t}=\frac{L}{E}$
(for Keplerian disk $\theta=\frac{\pi}{2}$) known as the projection of
impact parameter along the z-axis. So, like Luminet (1979), we consider the
final form of redshift as
\begin{equation}
1+z = (1-3/r)^{-1/2}[1+(1/r^{3/2})b\mathrm{sin}\theta_{\circ}\mathrm{sin}{\alpha}],
\end{equation}
where, $(1-3/r)^{-1/2}=z_{grav}$ and $(1+(1/r^{3/2})b\mathrm{sin}\theta_{\circ}\mathrm{sin}{\alpha})=z_{rot}$.

\section{Disk Radiation}
We use the time averaged radiation profile given by Page and Thorne (1974). 
Bolometric flux from a Keplerian disk is written as
\begin{equation}
F_{k}^{disk}(r) = \frac{F_c(\dot{M})}{(r-3/2)r^{5/2}}\bigg[\sqrt{r}-\sqrt{3}+\frac{\sqrt{3/2}}{2}log\bigg(\frac{(\sqrt{r}+\sqrt{3/2})(\sqrt{3}-\sqrt{3/2})}{(\sqrt{r}-\sqrt{3/2})(\sqrt{3}+\sqrt{3/2})}\bigg)\bigg],
\end{equation}

where, $F_c(\dot{M})=\frac{3M\dot{M}}{8\pi r_{g}^3}$, with $\dot{M}$ being the disk
accretion rate in Eddington unit. We keep $\dot{M}=0.1 M_{edd}$ throughout our simulation
for Keplerian disk. This is the flux emitted from source. Observed flux and emitted
flux relation is given by,
\begin{equation}
F_{k}^{obs}= \frac{F_{k}^{disk}(r)}{(1+z)^4}.
\end{equation}
Fourth power in the redshift factor comes due to the energy loss of photons (one power), 
gravitational time dilation (one power) and relativistic correction of detector solid angle
(two powers). The maximum observed flux is located
at $r(F_{k}^{max})=4.77r_g$ with $F_{k}^{max}\sim4\times10^{-4}$. From Stefan-Boltzmann
law, one can easily get observed temperature of disk by
\begin{equation}
T_{k}^{obs} = \bigg(\frac{F_{k}^{obs}}{\sigma}\bigg)^{1/4},
\end{equation}
where $\sigma=\frac{2\pi^5k^4}{15h^3c^3}$ is the Stefan-Boltzmann constant.

\def\figsubcap#1{\par\noindent\centering\footnotesize(#1)}
\begin{figure}[hold]%
\begin{center}    
    \parbox{1.5in}{\includegraphics[width=.34\textwidth,angle= 0]{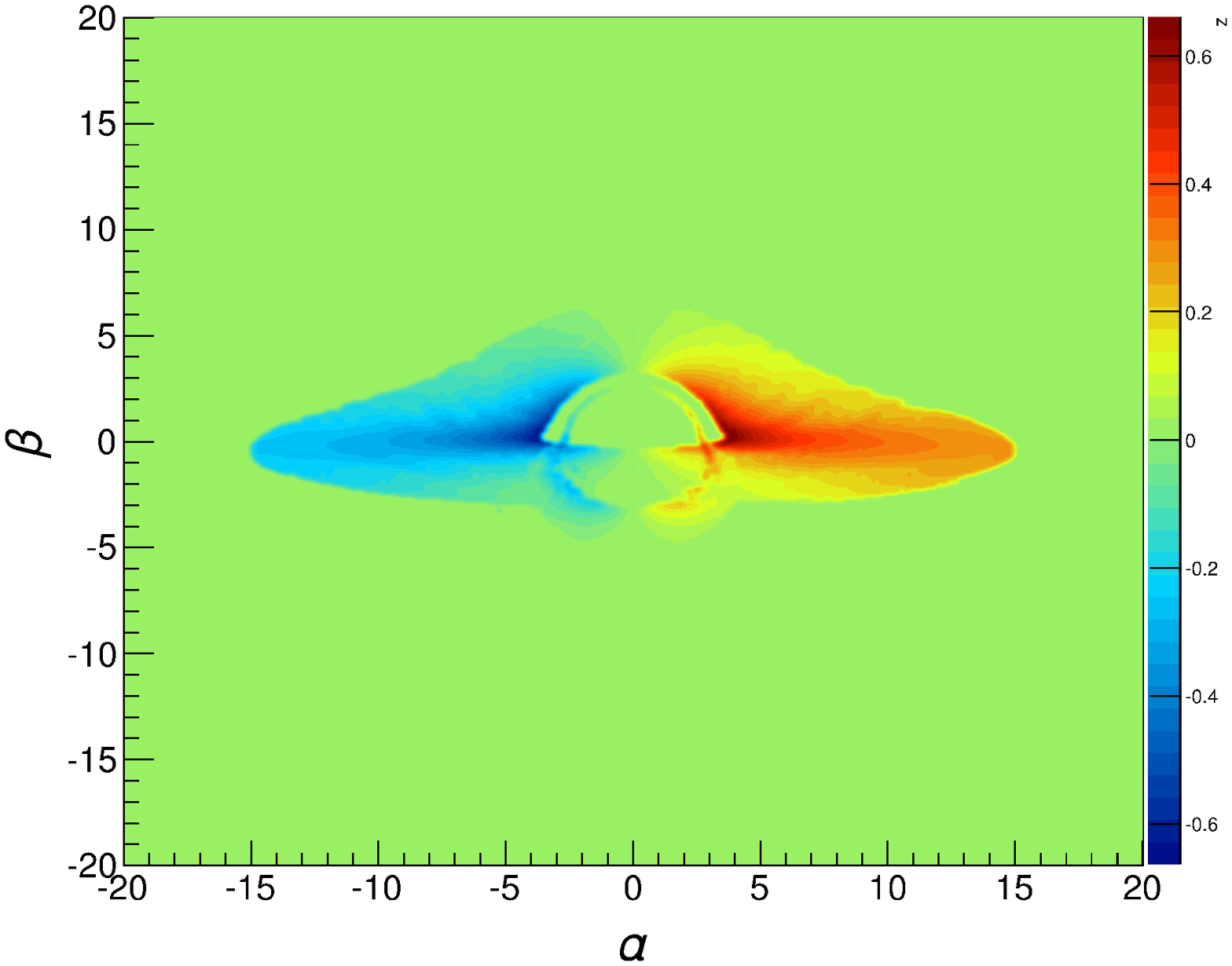}\figsubcap{a}}
    \hspace*{2pt}
    \parbox{1.5in}{\includegraphics[width=.34\textwidth,angle= 0]{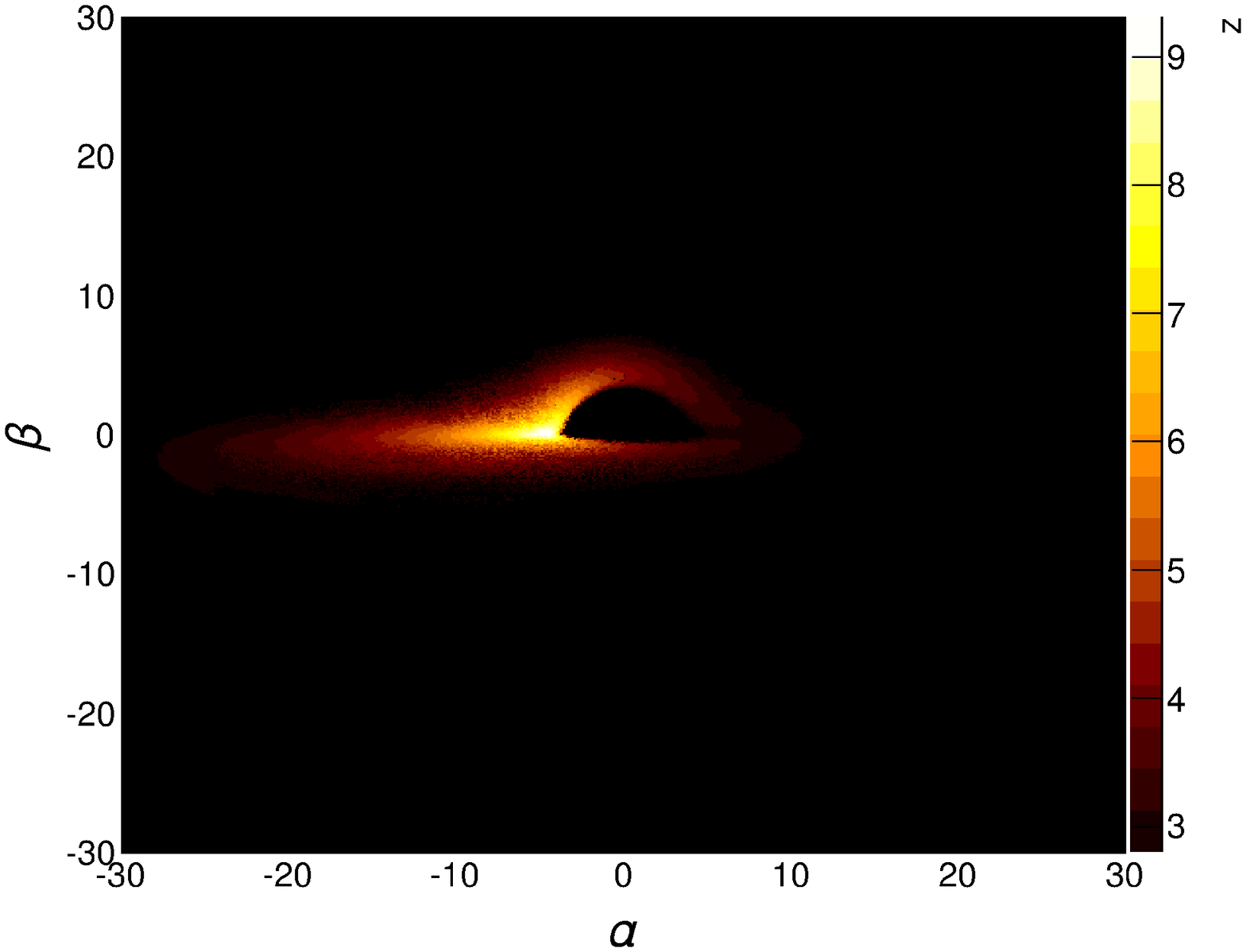}\figsubcap{b}}
    \hspace*{2pt}
    \parbox{1.5in}{\includegraphics[width=.34\textwidth,angle= 0]{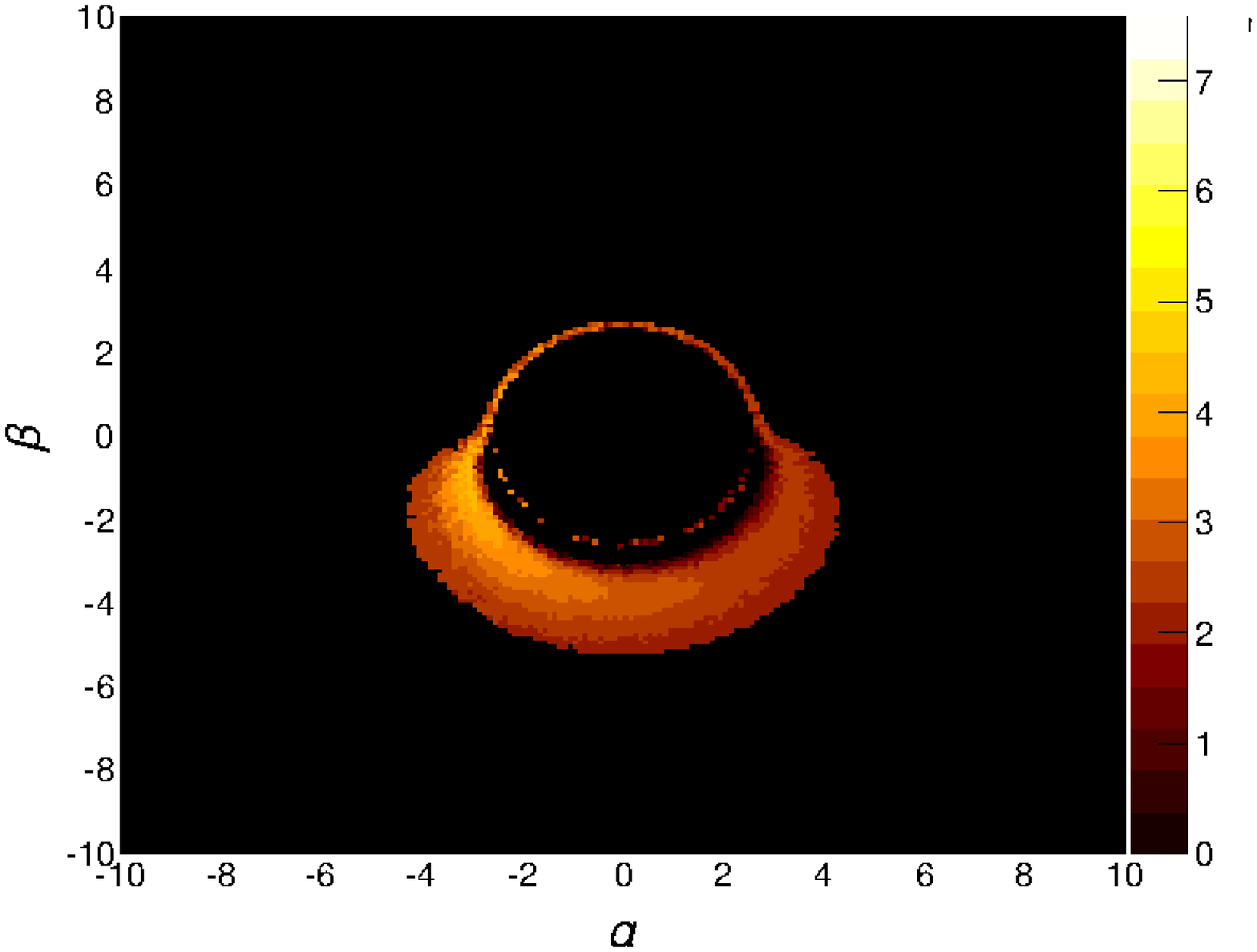}\figsubcap{c}}
        \caption{(a) Doppler tomography of the Keplerian Disk. All
         photons that are coming towards the observer have been considered in this 
         picture. (b) Keplerian disk seen at an inclination angle $\theta_{\circ}=82^{\circ}$ 
         (only direct photons are considered). (c) Shows all secondary (photons that have 
         come out after encircling the black hole) photons of different energy bands. 
         Color gradient represent normalized temperature.}
 \label{fig1.2}
\end{center}
\end{figure}

To correlate energy and temperature we used Wien's displacement law. Temperature is 
associated with the radius. So, the obtained most probable frequency will be related 
to the temperature via $\nu_{m}=5.89\times10^{10}T(r)~\mathrm{Hz}$.

\section{Thick Disks}
Basic process of generating thick disks involves a suitable solution 
of Euler equation which is given by 
\begin{equation}
\frac{\nabla p}{p+\epsilon} = -\ln(u_{t}) + \frac{\Omega \nabla l}{(1-\Omega l)},
\end{equation}
where, p is pressure and $\epsilon$ is total energy density. Also, $l=-u_{\phi}/u_{t}$ is 
specific angular momentum and $\Omega =u^{\phi}/u^{t}$ is the relativistic angular momentum.
If we choose barotropic process $p=p(\epsilon)$, then constant pressure surfaces and 
equipotential surfaces coincides. Under such condition, Euler equation becomes
\begin{equation}
W-W_{in} = \int_0^p\frac{dp}{p+\epsilon}=\int_{{u_{t}}_{in}}^{u_{t}} \ln(u_{t}) - \int_{l_{in}}^{l}\frac{\Omega \nabla l}{(1-\Omega l)},
\end{equation}
where, for a given von Zeipel relation, one can easily integrate the third term 
of this equation. Of course, making $l=constant$ is one easier solution. This 
type of distribution is effective for the open curves where matter falls onto 
black hole without transporting angular momentum. When advection is added, the
closed curves open up (Chakrabarti 1993). We use a generalized von Zeipel 
relationship, $\Omega=\Omega(l)$ from Chakrabarti (1985). 
The ratio $l/\Omega=\lambda^2$ has dimension $[L^2]$. Choosing 
$l=c\lambda^n$ (Chakrabarti 1985), where c and n are constants, yields the von Zeipel relationship as follows
\begin{equation}
\Omega = c^{2/n}l^{1-2/n}.
\end{equation}

\begin{figure}
\centering
\includegraphics[width=.4\textwidth,angle= 0]{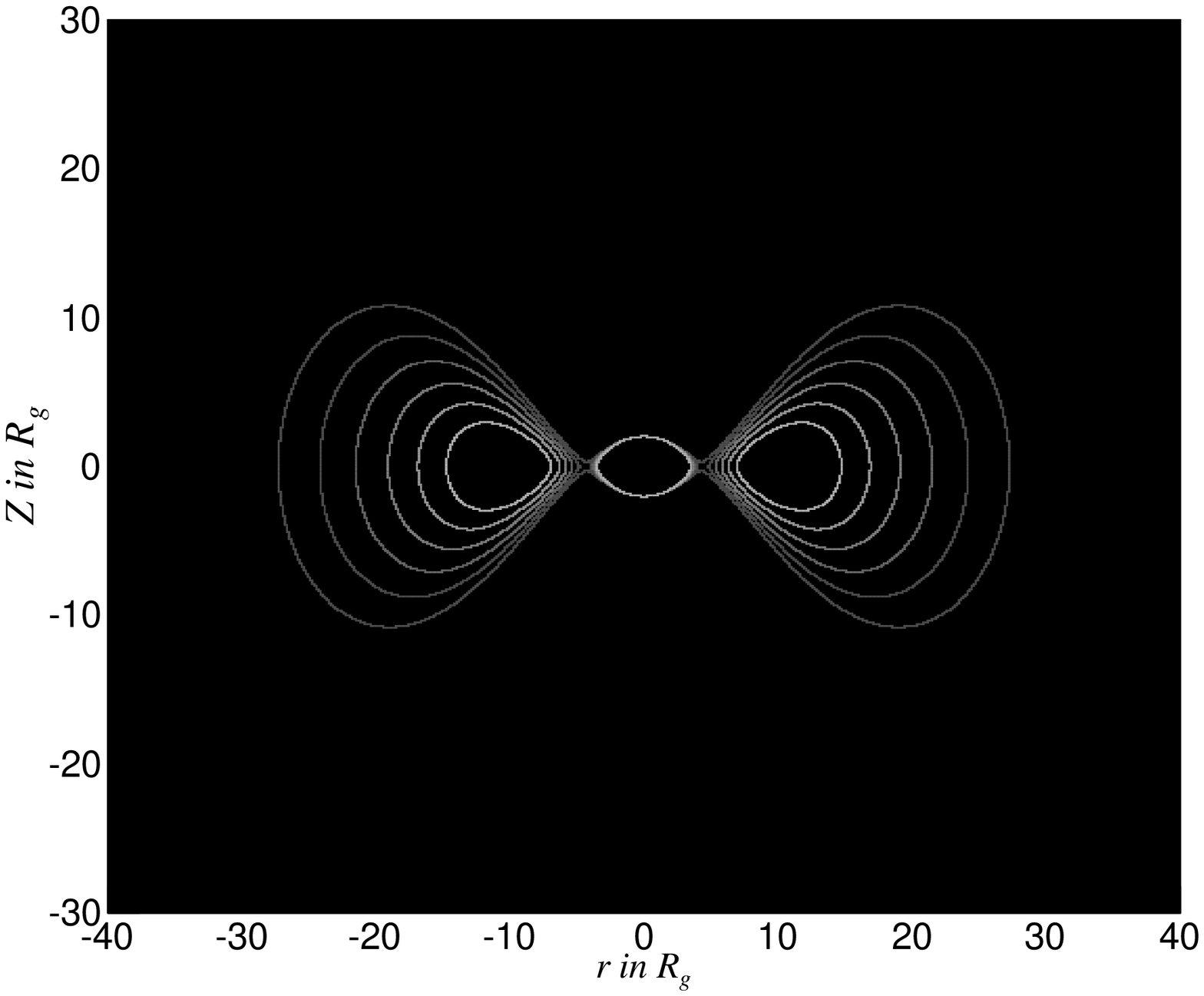}
\includegraphics[width=.4\textwidth,angle= 0]{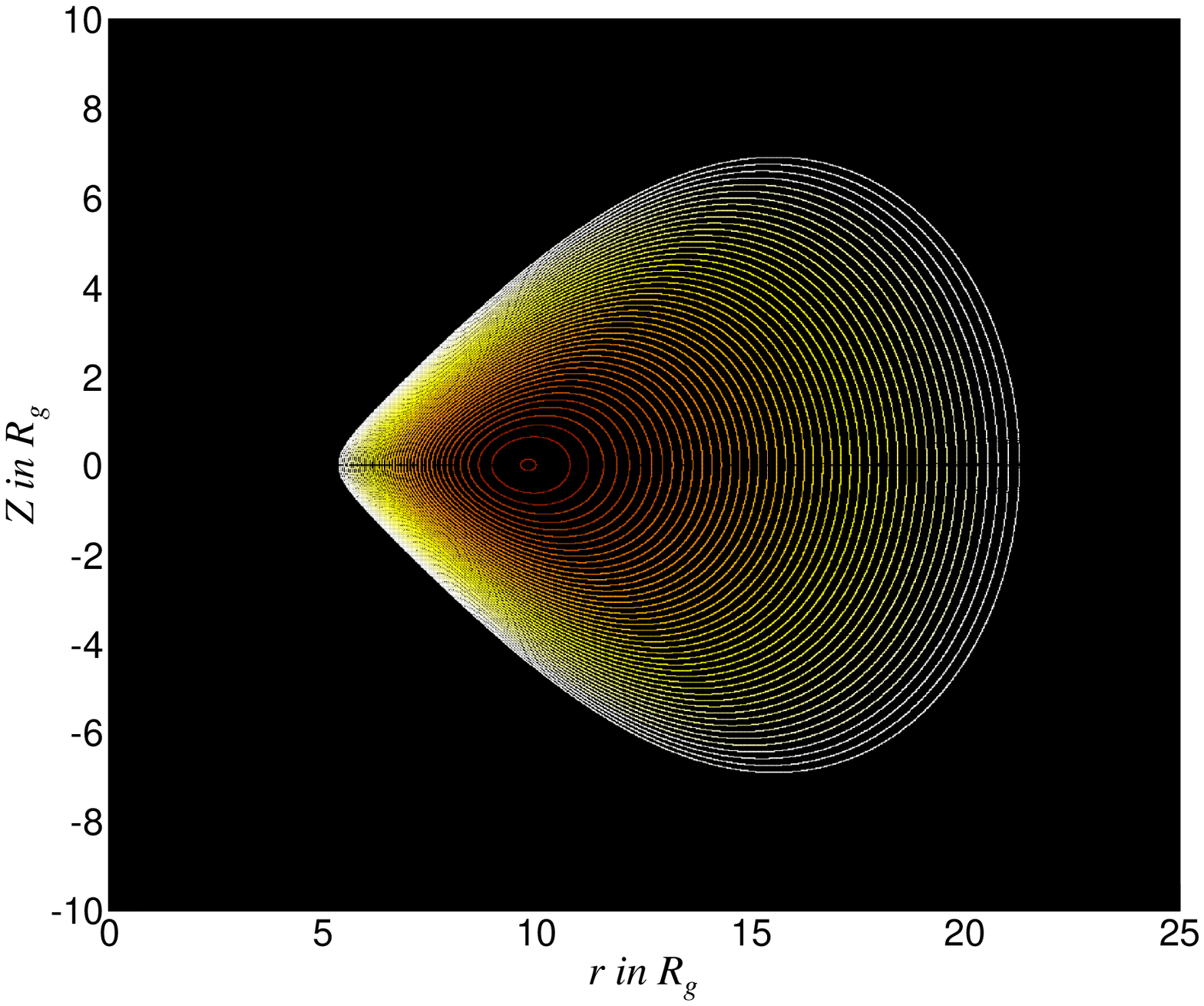}
  \caption{Different thick disk shells for $r_{in}=5.4~R_g$ and $r_{c}=9.8~R_g$. First figure
           contains open and closed curves. Second one corresponds to only closed equipotential
           surfaces.}
\end{figure}

From this, one can obtain all the variations such as 1. $ l=l_{0}=constant$ 
for $n=0$ (Abramowicz et al. 1978), 2. $l_{*}=\frac{l}{1-\Omega l}=constant$ for $n(r_{in})<0.268$ 
for constant c (Kozlowski et al. 1978) and many more by using different values of n. The rigid 
body rotation condition is satisfied using $n=2$. Constant parameters c and n can 
be derived from $c\lambda^{n}_{in} = l_{k}(r=r_{in},\theta=\pi/2)$~and~$c\lambda^{n}_{c} = l_{k}(r=r_{c},\theta=\pi/2),$
where, the subscripts $in$ and $c$ correspond for the disk inner edge and center of the 
disk. Thermodynamical properties, such as temperature, pressure and density inside the
thick disk were calculated by using Chakrabarti et al. (1987).

\section{Results and Conclusion}
We take Keplerian disk with uniform particle density. Radiation from disk 
particles are isotropic in nature. They are moving in time like circular orbits with 
no radial velocity. Keplerian angular velocity of disk particles is given as $\Omega_k =r^{-3/2}$.

\def\figsubcap#1{\par\noindent\centering\footnotesize(#1)}
\begin{figure}[!htb]%
\begin{center}
  \parbox{1.5in}{\includegraphics[width=3.5cm,height=3cm]{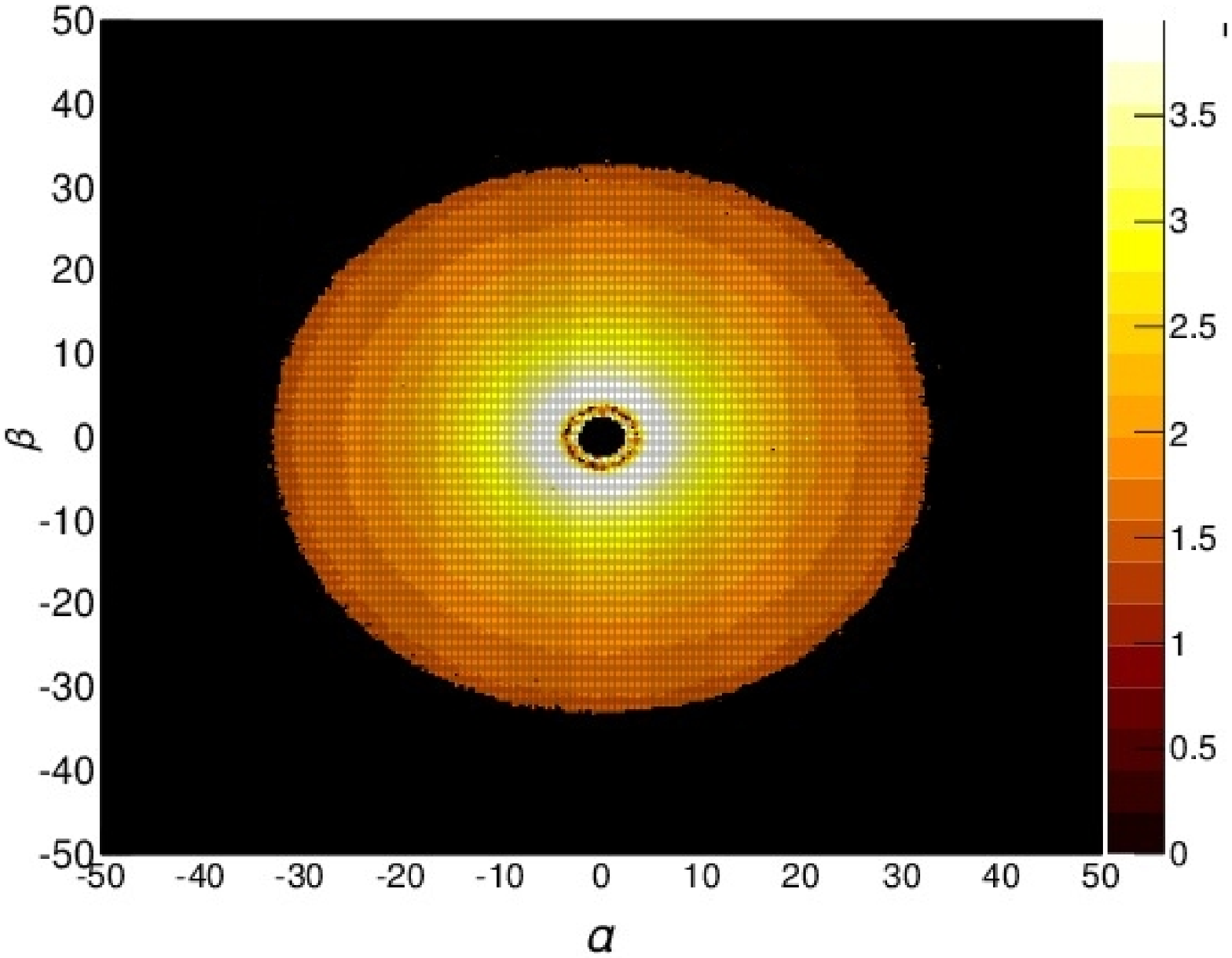}\figsubcap{a}}
  \hspace*{2pt}
  \parbox{1.5in}{\includegraphics[width=3.5cm,height=3cm]{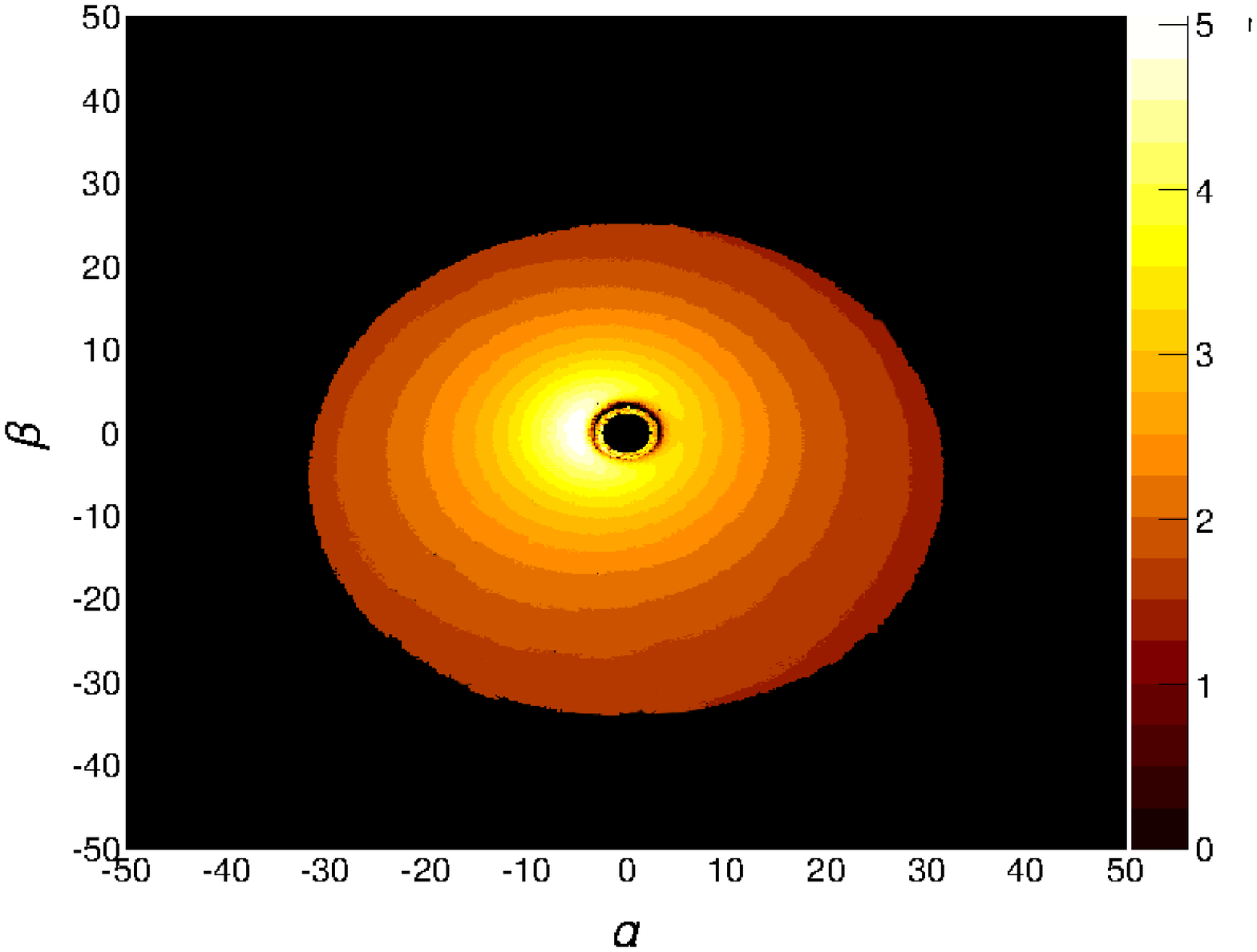}\figsubcap{b}}
  \hspace*{2pt}
  \parbox{1.5in}{\includegraphics[width=3.5cm,height=3cm]{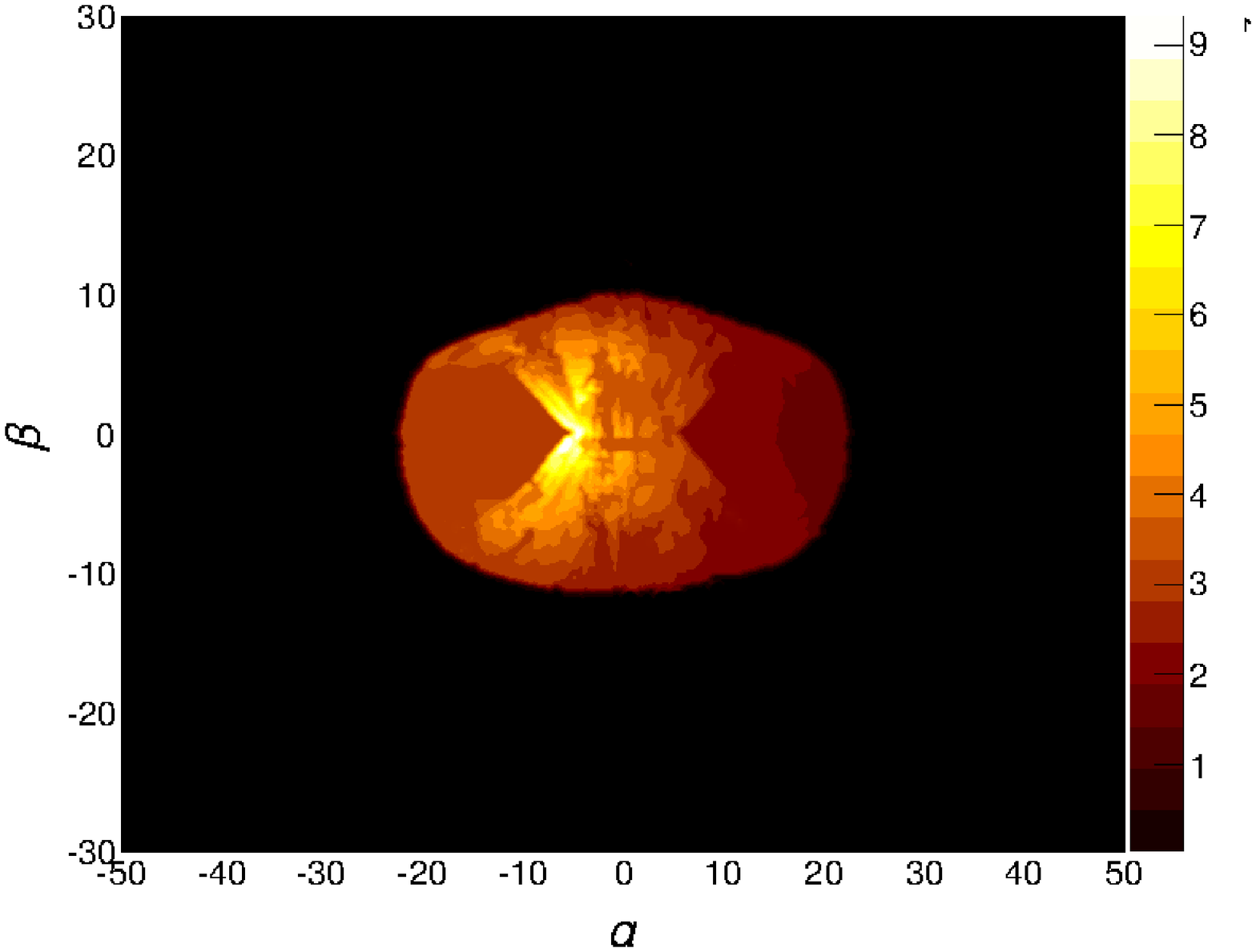}\figsubcap{c}}

  \parbox{1.5in}{\includegraphics[width=3.5cm,height=3cm]{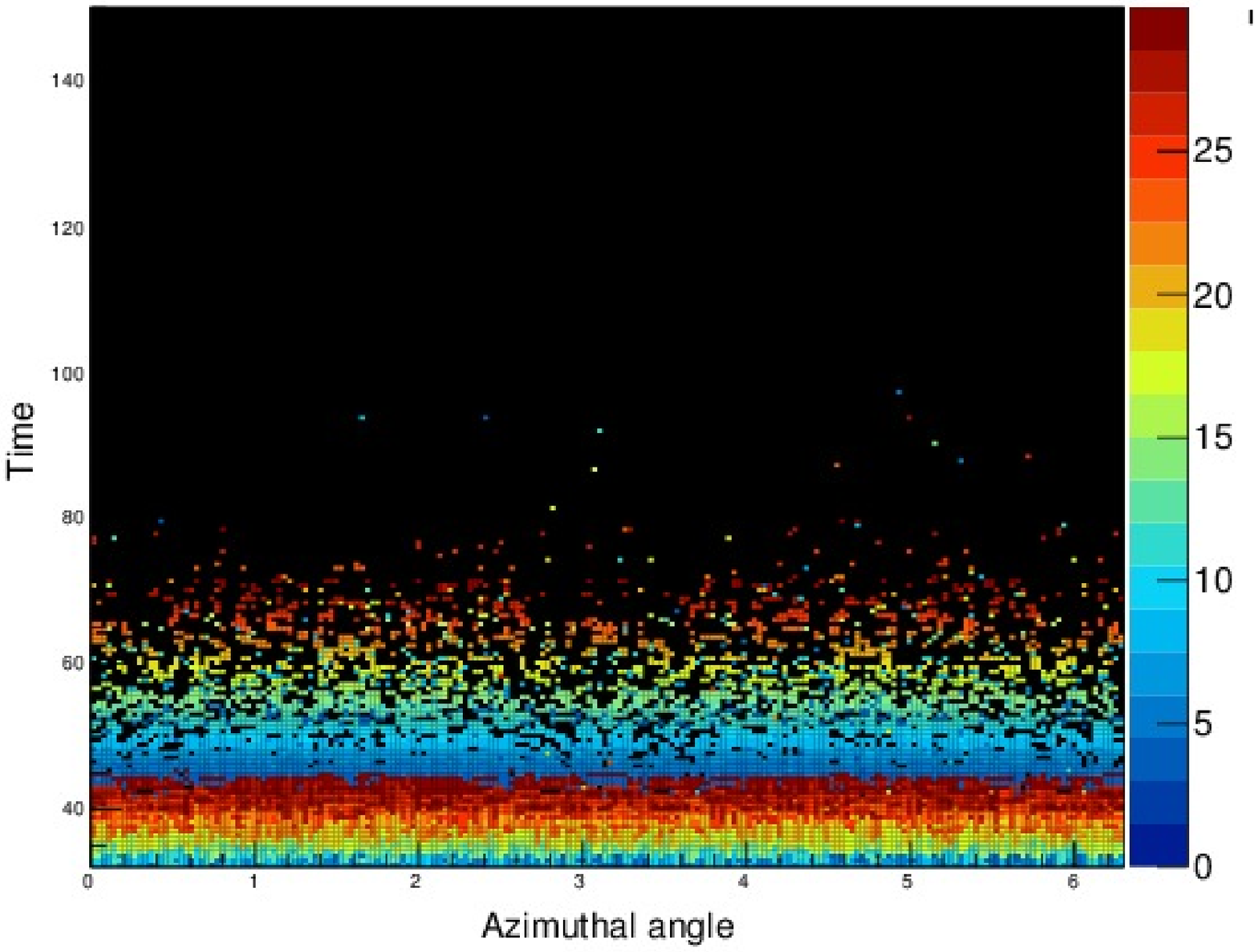}\figsubcap{d}}
  \hspace*{2pt}
  \parbox{1.5in}{\includegraphics[width=3.5cm,height=3cm]{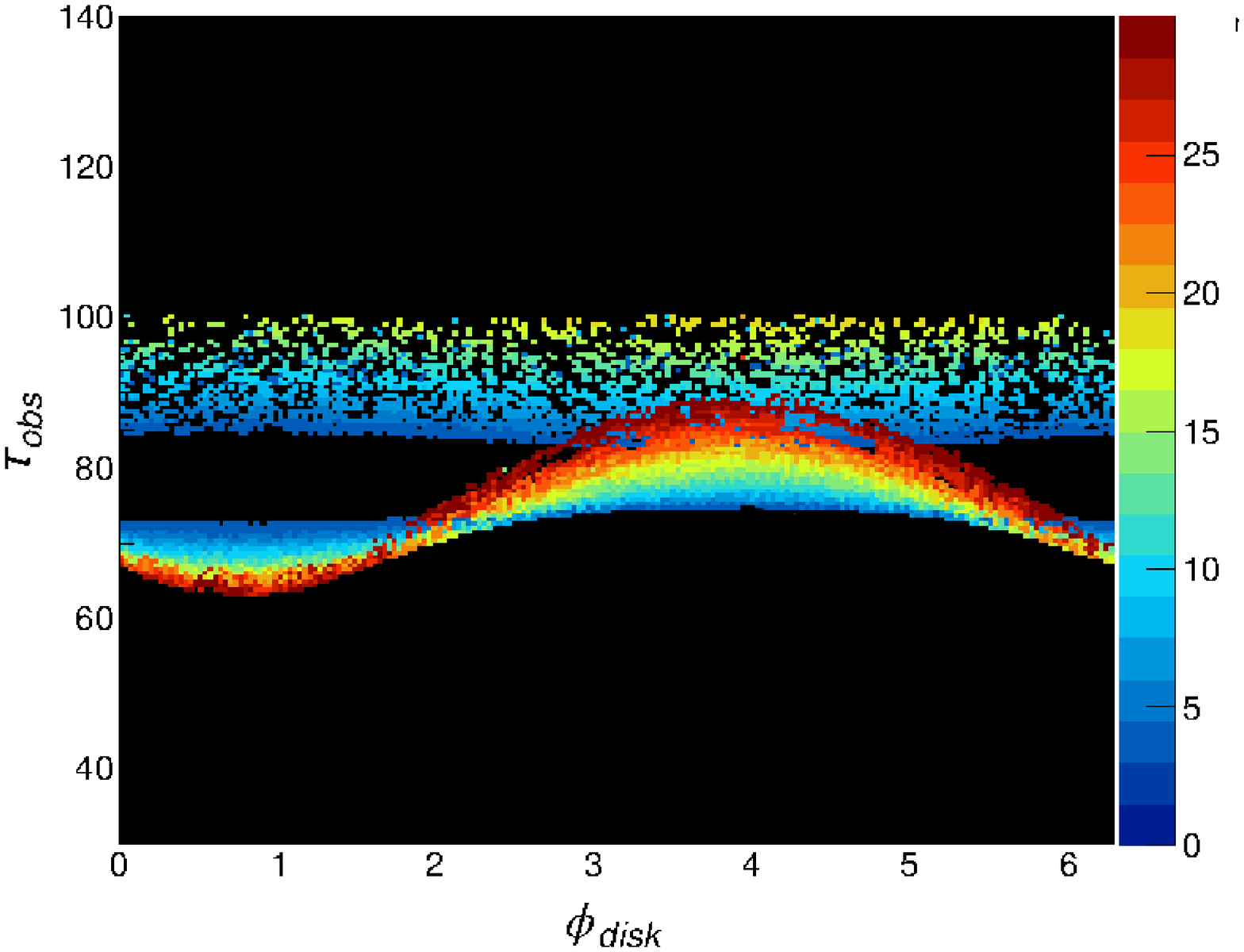}\figsubcap{e}}
  \hspace*{2pt}
  \parbox{1.5in}{\includegraphics[width=3.5cm,height=3cm]{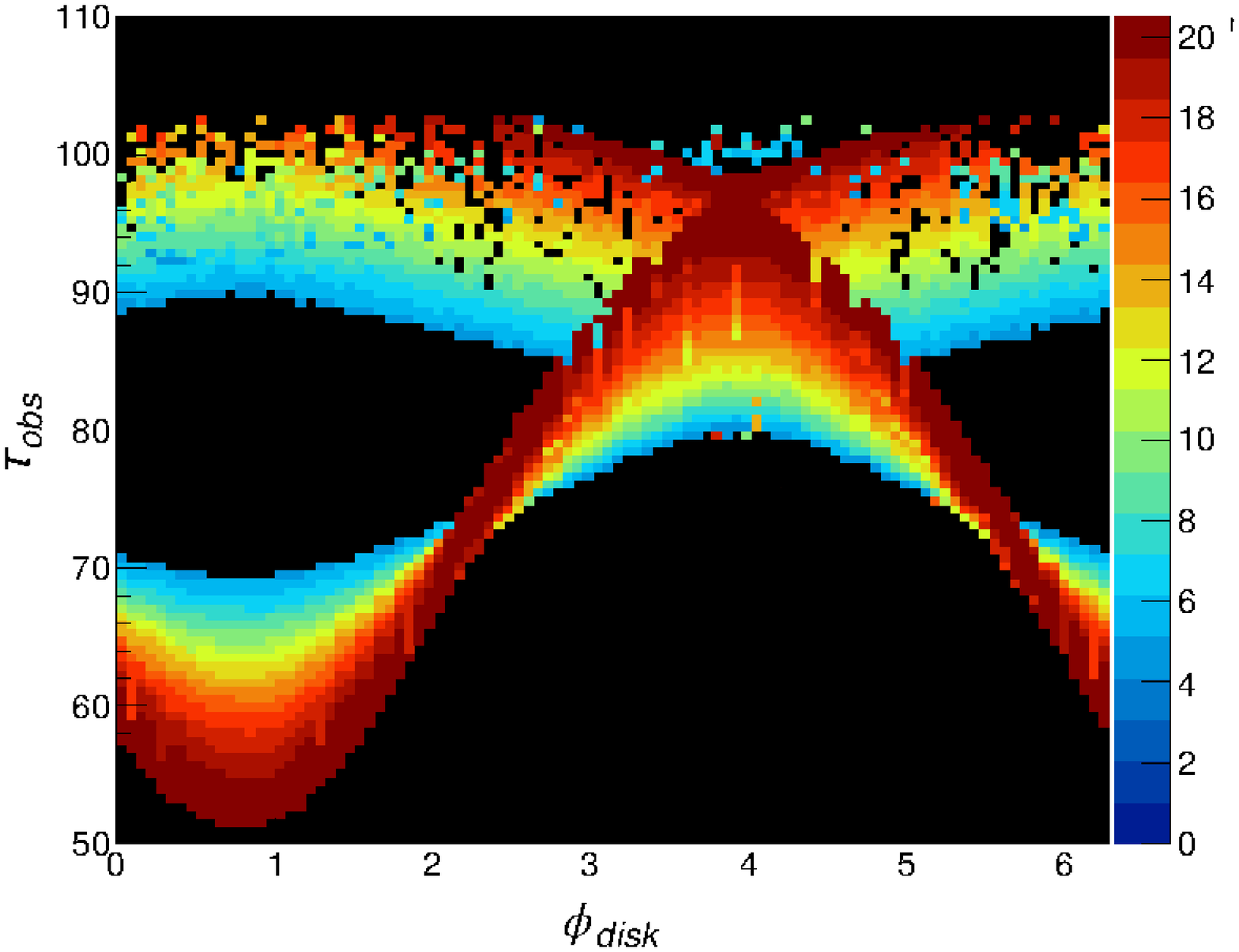}\figsubcap{f}}

  \parbox{1.5in}{\includegraphics[width=3.5cm,height=3cm]{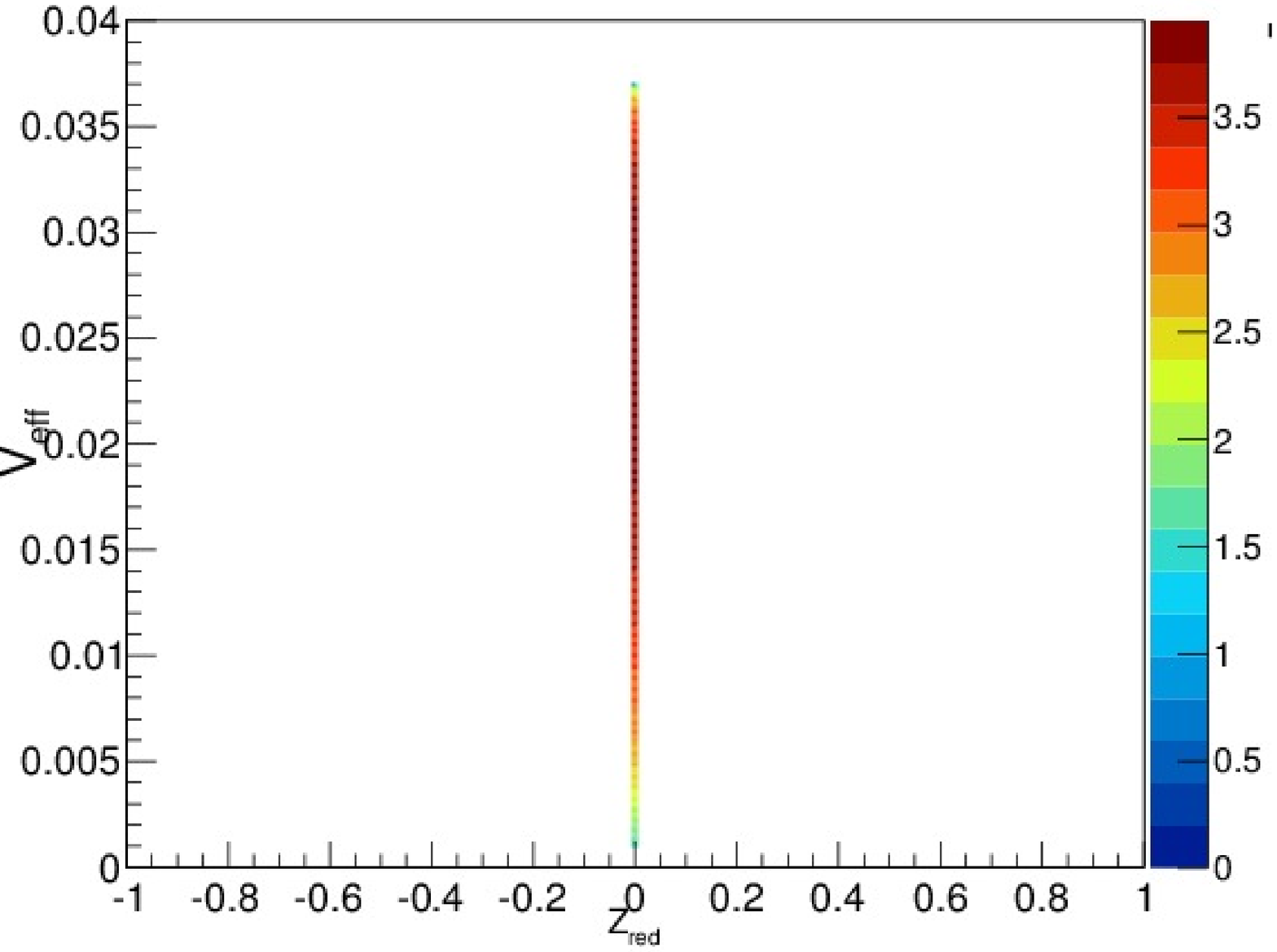}\figsubcap{g}}
  \hspace*{2pt}
  \parbox{1.5in}{\includegraphics[width=3.5cm,height=3cm]{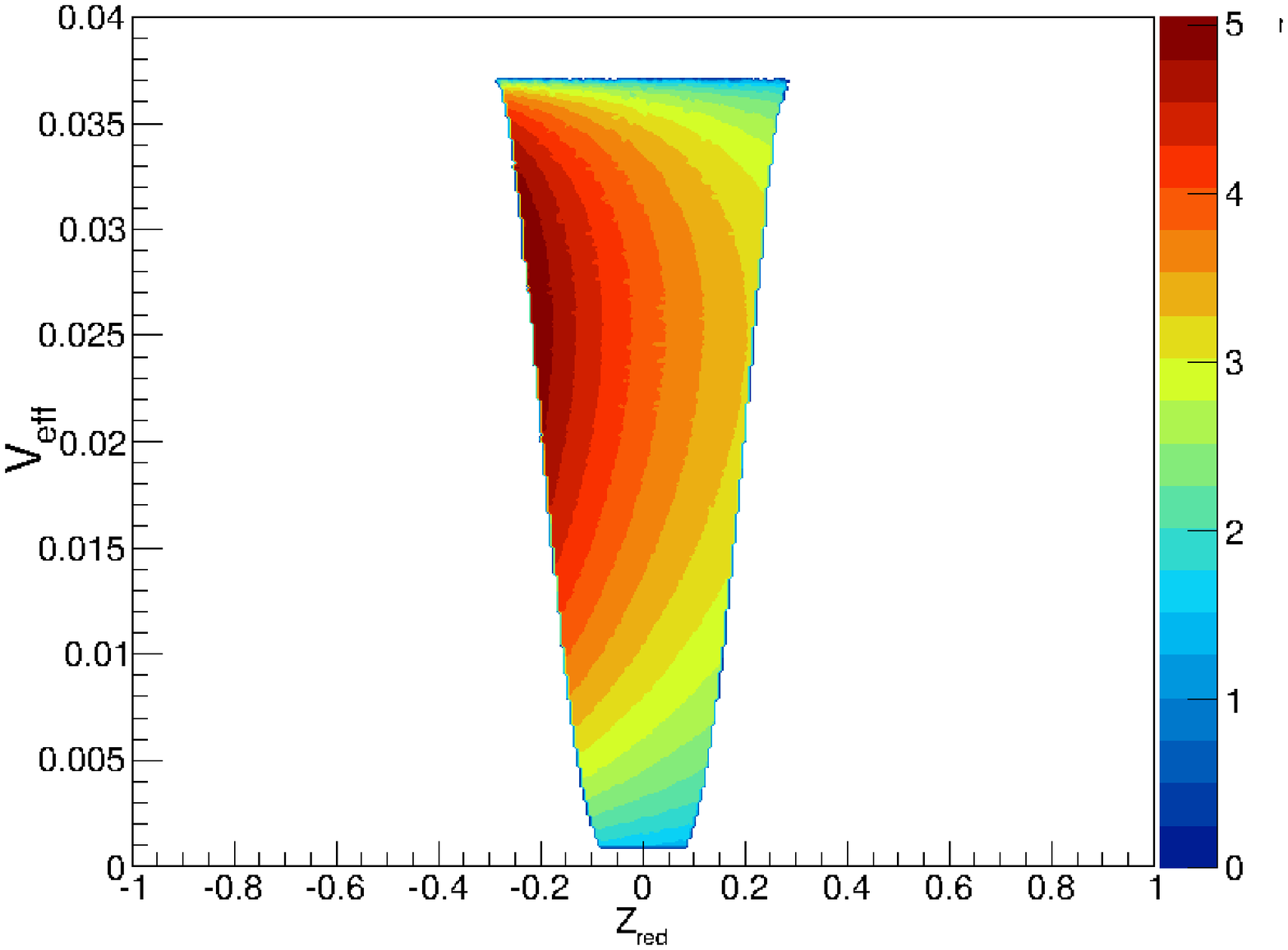}\figsubcap{h}}
  \hspace*{2pt}
  \parbox{1.5in}{\includegraphics[width=3.5cm,height=3cm]{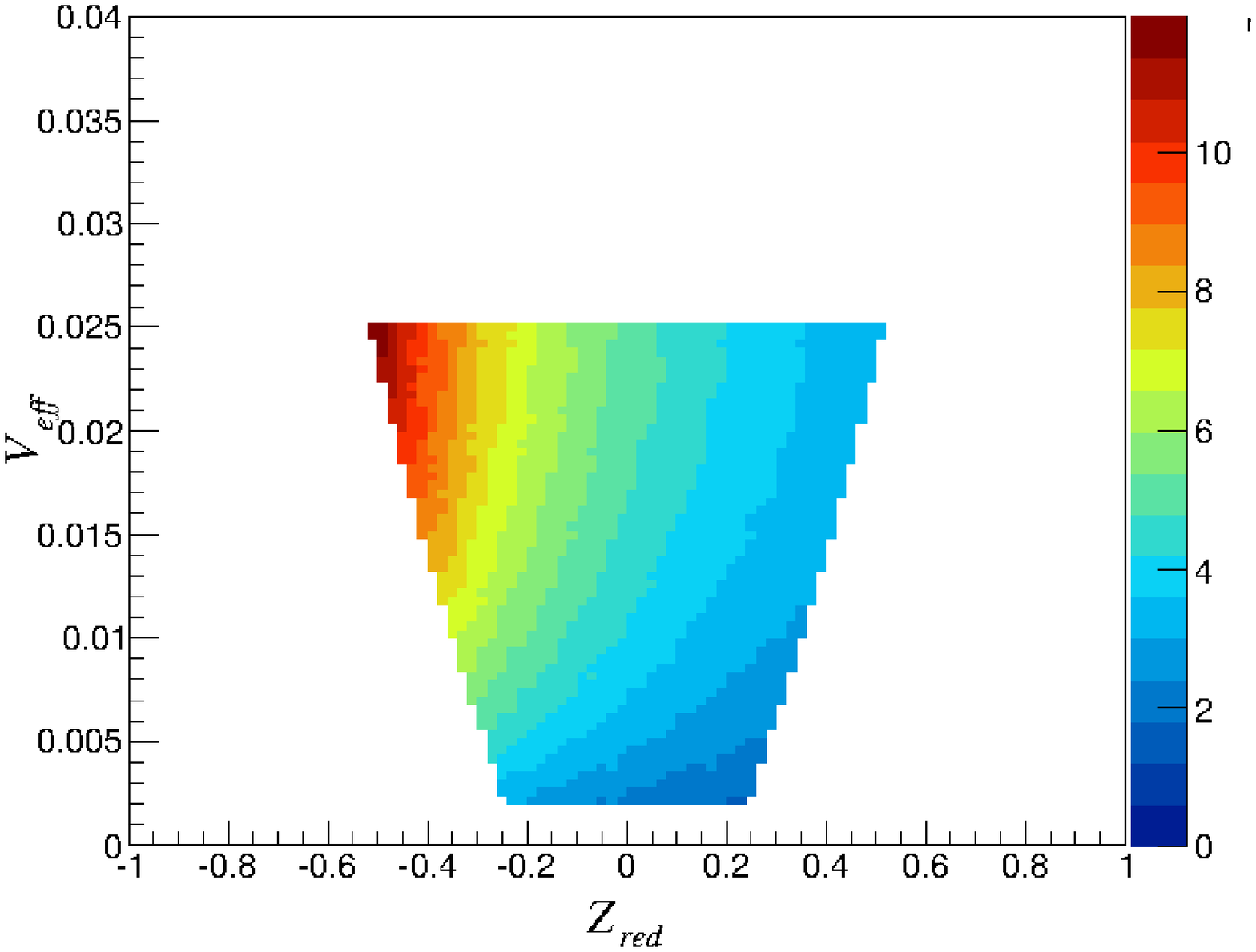}\figsubcap{i}}
  \caption{(a)-(b) Appearance of a Keplerian disk from $0.01^\circ$ \& $30^\circ$ 
inclination angles. A thick disk shell of $4.8-20.0$~$r_g$ where the inclination 
angle is $82^{\circ}$. (d), (e) \& (f) represent the corresponding time of arrival 
(in the unit of $2GM/c^3$) of photons to the observer from different azimuthal 
points on disk. Radial distance is the color bar here. (g), (h) \& (i) show 
parameter spaces consisting of Effective potential, Redshift and Temperature 
for $0.01^\circ$, $30^\circ$ \& $82^{\circ}$ respectively.}
\label{fig1.2}
\end{center}
\end{figure}

Observational results suggest that there are time lead or lag between high and low energy photons 
which led us to study time of arrival of each photon. We found a significant amount of 
time difference ($~6ms$ and higher) in higher inclination angle cases. Redshift produces 
more energetic photons at higher inclination angles. Also, the Doppler line broadening 
features increase with the increasing inclination angle.\\
We show a particular case of thick disk shell between $4.8-20$ $r_g$. In a TCAF we not 
only observe multicolor Blackbody radiation from the standard Keplerian disk component,
we also observe power law component by Comptonization of the photons intercepted by the 
thick disk. In future, we present these results, both continuum and line emission around 
Schwarzschild and Kerr black holes.

\section*{Acknowledgments}
The work of AC is supported by a grant from Ministry of Earth Science, Government of India.
AC acknowledges the MG14 conference organizers.

\end{document}